\documentclass[twocolumn]{pasj00}

\SetRunningHead{Yoshida et al.}{Two PL States of the ULX IC\,342 X-1}

\Received{$\langle$reception date$\rangle$}
\Accepted{$\langle$acception date$\rangle$}

\usepackage{times}
\usepackage{color}

\begin{document}

\title{Two Power-Law States of the Ultraluminous X-Ray Source IC\,342 X-1}

\author{
Tessei~\textsc{Yoshida}\altaffilmark{1},
Naoki~\textsc{Isobe}\altaffilmark{2},
Shin~\textsc{Mineshige}\altaffilmark{1},
Aya~\textsc{Kubota}\altaffilmark{3},
Tsunefumi~\textsc{Mizuno}\altaffilmark{4},
and Kei~\textsc{Saitou}\altaffilmark{2,5}
}

\altaffiltext{1}{Department of Astronomy, Kyoto University, Kitashirakawa-Oiwakecho, Sakyo, Kyoto 606-8502}
\altaffiltext{2}{Japan Aerospace Exploration Agency, Institute of Space and Astronautical Science, 3-1-1 Yoshinodai, Chuo, Sagamihara, Kanagawa 252-5210}
\altaffiltext{3}{Department of Electronic Information Systems, Shibaura Institute of Technology, 307 Fukasaku, Minuma, Saitama 337-8570}
\altaffiltext{4}{Department of Physics, Hiroshima University, 1-3-1 Kagamiyama, Higashi-Hiroshima, Hiroshima 739-8526}
\altaffiltext{5}{Department of Astronomy, The University of Tokyo, 7-3-1 Hongo, Bunkyo, Tokyo 113-0033}

\email{yoshida\_tessei@kusastro.kyoto-u.ac.jp}

\KeyWords{accretion, accretion disks --- black hole physics --- X-rays: individual (IC\,342 X-1)}

\maketitle

\begin{abstract}

In order to elucidate the emission properties of ultraluminous X-ray sources (ULXs) during their power-law (PL) state,
we examined long-term X-ray spectral data of IC\,342 X-1 during its PL state
by using our own Suzaku data and the archival data by XMM-Newton, Chandra, and Swift observations.
The PL state of this source seems to be classified into two sub-states in terms of the X-ray luminosities in 0.5--10~keV:
the low luminosity PL state with (4--6)$\times 10^{39}$~erg~s$^{-1}$ and the high luminosity one with (1.1--1.4)$\times 10^{40}$~erg~s$^{-1}$.
During the Suzaku observations which were made in 2010 August and 2011 March, X-1 stayed in the low luminosity PL state.
The observed X-ray luminosity (4.9--5.6$\times$10$^{39}$~erg~s$^{-1}$)
and the spectral shape (photon index~=~1.67--1.83) slightly changed between the two observations.
Using the Suzaku PIN detector,
we for the first time confirmed a PL tail extending up to at least 20~keV with no signatures of a high-energy turnover
in both of the Suzaku observations.
In contrast, a turnover at about 6~keV was observed during the high luminosity PL state in 2004 and 2005 with XMM-Newton.
Importantly, photon indices are similar between the two PL states and so is the Compton $y$-parameters of $y$~$\sim$~1,
which indicates a similar energy balance (between the corona and the accretion disk)
holding in the two PL states despite different electron temperatures.
From spectral similarities with recent studies about other ULXs and the Galactic black hole binary GRS\,1915$+$105,
IC\,342 X-1 is also likely to be in a state with a supercritical accretion rate,
although more sensitive higher energy observations would be necessary to conclude.
  
\end{abstract}

\section{Introduction}

Ultraluminous X-ray sources (ULXs; \cite{max00}) continue to be enigmatic objects since their discovery in the 1980's;
see \citet{fab04} and \citet{feng11} for reviews.
They are bright, variable, and off-nuclear point-like X-ray sources with luminosities of $>$10$^{39-41}$~erg~s$^{-1}$,
which exceed the Eddington luminosity ($L_{\rm{Edd}}$) of Galactic black holes (BHs).
Despite intensive multi-wavelength studies over three decades, their central engine remains an open question.
ULXs are very likely to contain a BH, but not a super-massive BH,
since their X-ray properties are similar to those of Galactic BH binaries (BHBs) in many respects;
e.g., they show short-term flux variations and have records of exhibiting at least two spectral states
characterized by power-law (PL) shaped spectra and convex-shaped (thermal) spectra \citep{max07,sor11}.
Some ULXs have even exhibited spectral transitions between the two \citep{laPar01,kubo01,iso09}.
These similarities indicate the same mechanism (i.e., gas accretion onto BHs) being responsible for electromagnetic radiation in ULXs.

Some authors (e.g., \cite{kubo02,mizu07,max07,iso09}) claimed that
the PL and the convex spectral states of ULXs may correspond to the very high state (VHS; \cite{miyamo91})
and an extremely high luminous disk state \citep{kubo04a,iso12} described by a slim disk model \citep{abr88} of Galactic BHBs, respectively.
This interpretation is based on the following important observational facts:
(i) Some observed properties of ULXs are consistent with the theoretical study based on the slim disk model \citep{wata00}.
For example, the derived temperature profile of the disk is flatter than that of the standard disk \citep{kiki06,tsuno06},
and the apparent innermost disk radius is inversely proportional to the innermost disk temperature \citep{mizu01}.
(ii) The luminosity difference between the two states of ULXs are consistent with that between the slim disk state and the VHS of Galactic BHBs.
Such ULXs were reported to be brighter in the convex state than in the PL state \citep{laPar01,kubo01,iso09}.

However, some concerns remain.
First, the PL state of ULXs shows a wider range of the photon indices;
for example, ${\it{\Gamma}}$~=~1.0--1.5 (M\,51 source-69 by \cite{yoshi10}),
1.6--2.0 (NGC\,1365 X-1 by \cite{sor09}), and 2.0--2.7 (NGC\,5204 X-1 by \cite{kaj09}).
Such various photon indices have not been reported in the VHS of Galactic BHBs.
For this concern, we are unable to rule out the possibility that the energy range of typical X-ray instruments (e.g., 0.5--10~keV)
corresponds to the different portion of the intrinsic X-ray spectrum in various types of ULXs (and BHBs),
and that it makes the variety in the photon index of ULXs.
The second concern is that a spectral turnover at 3.5--7~keV (e.g., \cite{kubo02,mizu07,gla09}), which is often accompanied with the PL spectra of ULXs,
is found at much lower energy than that of Galactic BHBs in the VHS (at about several tens~keV; \cite{kubo04b}).
These differences seem to imply that the PL state of ULXs and the VHS of Galactic BHBs could be distinct in nature.

\begin{figure*}
\begin{center}
\FigureFile(160mm,60mm){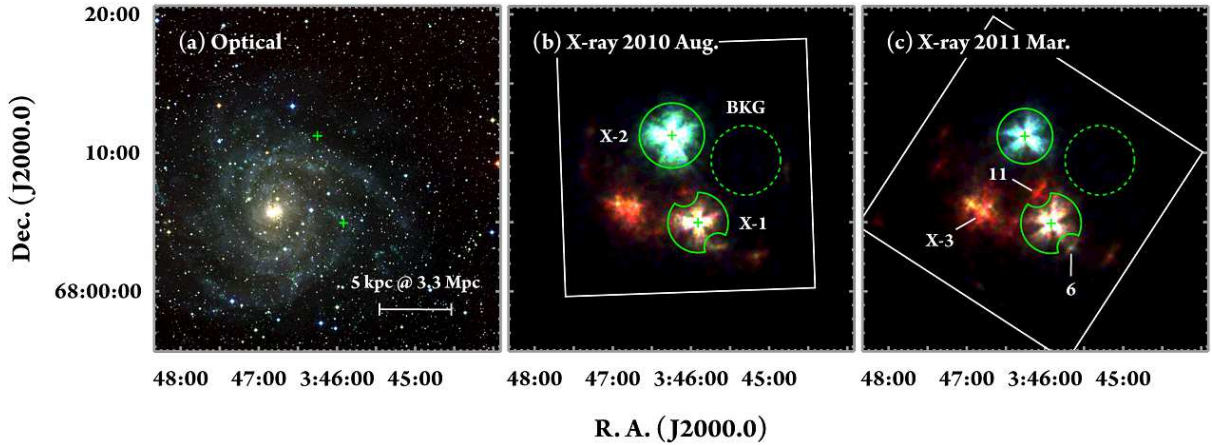}
\end{center}
\caption{Optical and X-ray images of IC\,342.
(a) Second Digitized Sky Survey three-color image.
Red, green, and blue represent the 850, 650, and 480~nm band intensities, respectively.
Pluses show the positions of the two ULXs.
(b)--(c) Smoothed Suzaku XIS three-color images obtained in (b) 2010 August and (c) 2011 March.
Red, green, and blue represent the 0.5--2, 2--4, and 4--10~keV bands, respectively.
Solid and dashed circles indicate the source and background extraction regions, respectively.
The other numbers are each source name (X-3, sources-6, and 11).
Solid squares show the FoV of XIS.
\label{fig1}}
\end{figure*}

Obviously, investigating the characteristics of ULXs in various PL states is a key to unveiling the origin of ULXs,
though they have not been so extensively discussed so far, compared with those in the convex state.
The suitable target to achieving our purpose is the unique ULX X-1 in the nearby spiral galaxy IC\,342 (figure~\ref{fig1}a) at 3.3~Mpc \citep{saha02}.
While this is the first ULX exhibiting the transition from the PL state to the convex one \citep{kubo01,kubo02},
it stayed in the PL state in most of the previous observations \citep{rob04,feng09,mak11}.
Further two seemingly distinct sub-states have been reported;
the high luminosity PL state and the low luminosity PL state,
where the former is brighter than the latter by a factor of 2 to 3 \citep{feng09,mak11}.

In order to elucidate the nature of the PL state of the source,
we discuss the PL spectral states and the long-term spectral variabilities of IC\,342 X-1 assembling all the recent available data sets,
which include our new data taken with Suzaku and unpublished data taken with Swift.
A particular attention is paid to the Suzaku data,
since it has a detector sensitive above 10~keV (see $\S$~2.1 for details),
thereby being able to determine the existence of a turnover at a hard X-ray band in the PL state.

\medskip

This paper is composed of the following sections:
In $\S$~2, we summarize the observations and data reduction.
In $\S$~3 and $\S$~4, we describe the results of timing and spectral analysis, respectively.
We develop discussion for the PL states of IC\,342 X-1 in $\S$~5, and conclude this paper in $\S$~6.

\section{Observations and Data Reduction}

Table~\ref{table1} shows the assembled data sets.
We labeled the observations as Su1--2 for Suzaku, XM1--4 for XMM-Newton, Ch1--2 for Chandra, and Sw1--3 for Swift.
The list includes all the available data sets except for a Chandra observation (ObsID~=~7069),
because the target source suffers a severe pile-up with a pile-up fraction of more than 20\%.
Throughout this paper, we use events in the 14--20~keV band for Suzaku PIN, and in 0.5--10~keV for the other instruments.

\subsection{Suzaku}

The Suzaku observatory \citep{mitsu07} has two operating science instruments.
One is the X-ray Imaging Spectrometer (XIS; \cite{koya07}) covering an energy range of 0.2--12~keV,
while the other is the Hard X-ray Detector (HXD; \cite{taka07,koku07}) covering 10--600~keV.

We observed IC\,342 with Suzaku twice on 2010 August (Su1) and 2011 March (Su2) for about 70~ks each (table~\ref{table1}).
We aimed at the middle point of X-1 and the other ULX (X-2) in order to observe the two ULXs within the XIS field of view (FoV).
The data in these observations were taken with the normal clocking mode with a frame time of 8~s for XIS,
and with the normal mode with the time resolution of 61~$\mu$s for HXD.

In order to create new cleaned event files,
we reprocessed the data using
the software package HEASoft\footnote{See http://heasarc.gsfc.nasa.gov/docs/software/lheasoft/ for details.} version~6.10
with the X-ray telescope calibration database (CALDB) version~20100730, the XIS CALDB version~20110210, and the HXD CALDB version~20101202.
The data were screened out during the South Atlantic Anomaly passages (SAA), within 436~s after exiting from the SAA,
with the Earth elevation angle of below 5\arcdeg, with the day-Earth elevation angle of below 20\arcdeg\ (for XIS),
or at low cutoff rigidity regions ($\le$6~GV for HXD).

\begin{table}
\begin{center}
\caption{Observation log.\label{table1}}
\footnotesize
\begin{tabular}{ccccc}
\hline
Data  & Observatory & ObsID       & Date       & $t_{\rm{exp}}$\footnotemark[$*$] \\
label &             &             &            & (ks) \\
\hline
Su1   & Suzaku     & 705009010   & 2010-08-07 & 74.4/64.1 \\
Su2   &            & 705009020   & 2011-03-20 & 68.5/74.3\\
XM1   & XMM-Newton & 0093640901  & 2001-02-11 & \phantom{1}9.7/\phantom{1}5.1 \\
XM2   &            & 0206890101  & 2004-02-20 & 23.4/17.8 \\
XM3   &            & 0206890201  & 2004-08-17 & 23.4/17.1 \\
XM4   &            & 0206890401  & 2005-02-10 & 18.6/15.1 \\
Ch1   & Chandra    & 2916        & 2002-04-29 & 9.3 \\
Ch2   &            & 2917        & 2002-08-26 & 9.9 \\
Sw1   & Swift      & 00035474001 & 2006-01-07 & 7.2 \\
Sw2   &            & 00031987001 & 2011-06-14 & 3.0 \\
Sw3   &            & 00031987002 & 2011-06-15 & 9.4 \\
\hline
\multicolumn{5}{@{}l@{}}{\hbox to 0pt{\parbox{85mm}{\footnotesize
\par\noindent
\footnotemark[$*$] XIS (left) and HXD PIN (right) exposures for Suzaku, EPIC MOS (left) and EPIC pn (right) exposures for XMM-Newton, ACIS exposure for Chandra, and XRT exposure for Swift.
}\hss}}
\end{tabular}
\end{center}
\end{table}

\medskip

\subsubsection{XIS}

XIS is composed of three front-illuminated (FI) and one back-illuminated (BI) CCD devices with a FoV of 17\farcm8$\times$17\farcm8.
Since one of the three FI devices became dysfunctional due to a putative micrometeorite hit in 2006 November,
we used the remaining two FI and one BI devices for our analysis.
We merged the two FI data because the devices have nearly identical responses.

Figures~\ref{fig1}~(b) and (c) show Suzaku XIS three-color images of IC\,342.
We can see the three bright sources (X-1, X-2, and X-3 named by \cite{oka98})
and two relatively faint sources (sources-6 and 11; \cite{bau03}).
X-2 and X-3 are respectively harder and softer than X-1 in the XIS range.
Since hard sources may contribute to the PIN spectrum (see $\S$~2.1.2), we also investigated the properties of X-2 in the present paper.
In the equinox J2000.0,
the peak positions of X-1 and X-2 were respectively determined at
(R.A., Dec.)~=~(\timeform{03h45m55s}, \timeform{+68D04'59''}) and (\timeform{03h46m15s}, \timeform{+68D11'17''}),
which are consistent with previous studies \citep{kubo01,kong03,bau03,mak11}.

Source events were extracted from a circle of an adaptively chosen radius of
130\arcsec, 130\arcsec, 140\arcsec, and 120\arcsec, for X-1 in Su1, X-1 in Su2, X-2 in Su1, and X-2 in Su2, respectively (figures~\ref{fig1}b and c).
Whereas, background events were extracted from a common region.
Here, we masked circles with a radius of 1\arcmin\ around the two relatively faint sources-6 and 11.
Other two relatively faint sources near each ULX (source-9 for X-1 and source-12 for X-2; \cite{kong03,bau03}) are not resolved in the XIS images,
but make no signification contamination.
Source-9 has 10\% of the X-ray flux of X-1, while source-12 has 3\% of X-2.
For the spectral analysis, we selected only the events with the standard ASCA grade set of 0, 2, 3, 4, and 6.

We calculated their redistribution matrix functions (RMFs) using the {\texttt{xisrmfgen}} tool.
Whereas, the ancillary response files (ARFs) were simulated by the {\texttt{xissimarfgen}} tool \citep{ishi07}.
The influence by relatively faint sources was removed.

\medskip

\subsubsection{HXD PIN}

HXD consists of two components:
PIN and GSO respectively sensitive to 10--70~keV and 40--600~keV X-rays.
In this paper, we focus on the PIN detector only, because the object is too faint for a GSO detection.
PIN is a non-imaging detector, which has a full width at zero intensity (FWZI) view of $\sim$70\arcmin\ square.
The effective area monotonically decreases as the distance increases from the field center.

If PIN has a high temperature, the data are affected by increased noise.
In order to deal with this situation, we further screened PIN events by applying the temperature threshold for observations in the epoch~10
(2010 December 1 to 2011 May 24)\footnote{See http://heasarc.nasa.gov/docs/suzaku/analysis/pinepochs.html for details.}.
The threshold excludes a noisy energy band of each PIN unit.
We used the detector response files distributed by the instrument team.

\subsection{XMM-Newton}

The details of the instruments and the observations can be found in \citet{feng09}, \citet{kaj09}, and \citet{mak11}.
We used the Science Analysis System (SAS) version~11.0.0 \citep{gab04} for reprocessing the data, extracting events, and generating response files.
We used the calibration files latest as of 2011 April.
We excluded the intervals with a high background rate,
which is defined as the 10--15~keV count rate of the entire array larger than the average by more than 3$\sigma$.
The events in the energy range are dominated by the background \citep{read03}.

A common background region devoid of bright X-ray sources was adopted.
We extracted source events from a circle of adaptively chosen radius of 45\arcsec\ for XM1, 55\arcsec\ for XM2, 55\arcsec\ for XM3, and 65\arcsec\ for XM4.
For the spectral analysis,
we selected the MOS events with PATTERN~$\le$~12, \#XMMEA\_EM, and FLAG~$=$~0, and the pn events with PATTERN~$\le$~4, \#XMMEA\_EP, and FLAG~$=$~0.
The RMF and ARF were generated with the SAS tools {\texttt{rmfgen}} and {\texttt{arfgen}}, respectively.

\subsection{Chandra}

The details of the instrument and the observations can be found in \citet{rob04}.
We reduced the data using the Chandra Interactive Analysis of Observations (CIAO) version~4.3 with the CALDB version~4.4.2.
We then extracted the events (with grade 0, 2, 3, 4, and 6),
and constructed the energy spectra using the ACIS Extract package \citep{bro02} version~2009-12-01.
The source events were accumulated from a region around each source encircling 90\% of photons of a point-like source.
The background events were from an annulus around each source.
The RMF and ARF were generated with the CIAO tools {\texttt{mkrmf}} and {\texttt{mkarf}}, respectively.

\subsection{Swift}

The Swift observatory \citep{geh04} is equipped with some instruments.
One of them is the X-ray telescope (XRT; \cite{bur05}) with an X-ray CCD device sensitive at 0.2--10~keV.
The FoV (23\farcm6$\times$23\farcm6) covers the whole of IC\,342.
In the observations, the photon counting mode with a time resolution of 2.5~s was used.

We processed the data sets based on the CALDB version~20110513 using HEASoft version~6.10.
In order to increase the photon statistics, two data sets (Sw2 and Sw3) with very close observation dates were merged.
Hereafter we call the merged observation Sw2$+$3.
Source and background regions were determined by the method same as for XMM-Newton,
and we then extracted the spectra from grade 0--12.
We used the RMF file in the CALDB, and we created the ARF using the {\texttt{xrtmkarf}} tool.

\section{Timing Analysis and Results}

\begin{figure}
\begin{center}
\FigureFile(68mm,107mm){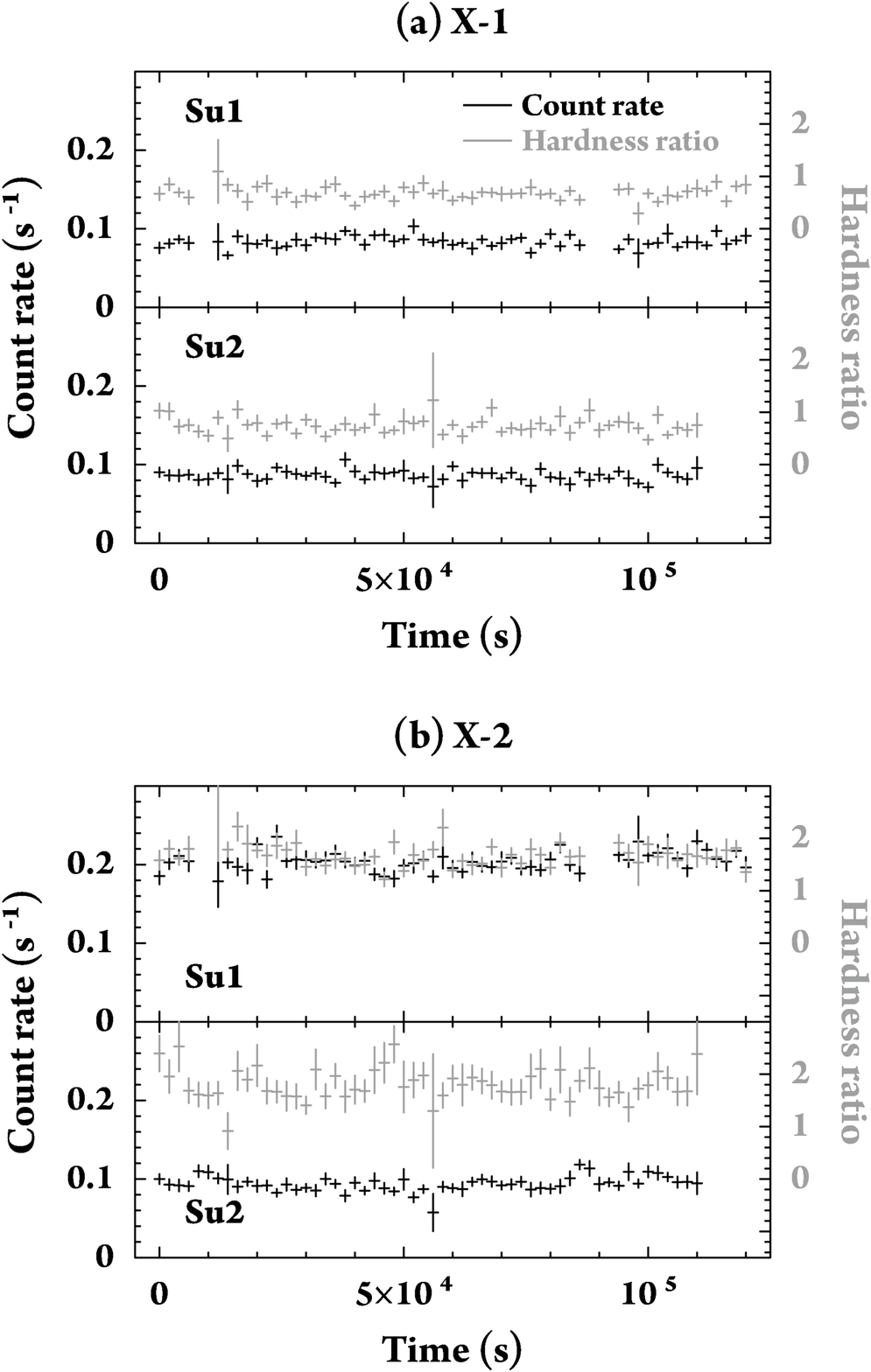}
\end{center}
\caption{Background-subtracted XIS FI light curves (black) and hardness ratio variations (gray)
of the two ULXs in 0.5--10~keV.
The curves are binned with 2000~s~bin$^{-1}$.
The hardness is calculated as the count rate ratio of 3--10~keV against 0.5--3~keV.
\label{fig2}}
\end{figure}

Figure~\ref{fig2} shows the background-subtracted Suzaku XIS (FI) light curves and
the hardness ratio (the 3.0--10.0~keV band count rate to the 0.5--3.0~keV band one) variations for each ULX.
No significant short-term variation was found in the light curves nor the hardness ratio variations at 99.9\% significance,
indicating that the spectra did not change within the individual exposure.
We thus stacked all the events in each observation to construct spectra.

For X-1, the time-averaged count rate and hardness ratio changed at most by several percent between the two Suzaku observations.
The time-averaged count rate was measured to be 0.083$\pm$0.001~counts~s$^{-1}$ (Su1) and 0.086$\pm$0.001~counts~s$^{-1}$ (Su2),
while the time-averaged harness ratio was 0.66$\pm$0.02 (Su1) and 0.71$\pm$0.02 (Su2).
For X-2, the count rate became to half, from 0.203$\pm$0.002~counts~s$^{-1}$ (Su1) to 0.093$\pm$0.001~counts~s$^{-1}$ (Su2).
Its hardness ratio (1.63$\pm$0.03 for Su1 and 1.74$\pm$0.05 for Su2) also changed only by several percent like X-1,
but was clearly higher than X-1 as mentioned from the three-color images (figure~\ref{fig1}).
Here, the errors are the 1$\sigma$ values.

The other data sets with XMM-Newton, Chandra, and Swift exhibit no short-term variation in each exposure.
We thus stacked all the events in each observation to construct spectra.

\section{Spectral Analysis and Results}

First, we explain spectral fitting models in $\S$~4.1.
Next, we report our Suzaku results based on the following three steps:
In $\S$~4.2.1, we fit the spectral models to the XIS spectra of X-1 and X-2, and describe the results.
Note that the spectra of X-2 are analyzed in order to examine the contribution to the PIN spectrum.
In $\S$~4.2.2, we estimate the source signals above 10~keV with the PIN detector by inspecting the NXB accuracy.
In $\S$~4.2.3, the XIS and PIN spectra are jointly analyzed.
Finally, we present the fitting results of the other satellites in $\S$~4.3.

\begin{table*}
\begin{center}
\caption{Best-fit parameters for the Suzaku XIS spectra of the two ULXs.\label{table2}}
\small
\begin{tabular}{cccccccc}
\hline
Name & Data  & Model & $N^{\prime}_{\rm{H}}$      & ${\it{\Gamma}}$ or $\tau$ & $E_{\rm{cut}}$, $k T_{\rm{e}}$, or $k T_{\rm{in}}$ & Luminosity\footnotemark[$*$]     & Red--$\chi^2$ \\
     & label &       & $(10^{22}\;\rm{cm}^{-2})$ &                           & (keV)                                           & $(10^{39}\;\rm{erg \; s^{-1}})$ & (d.o.f.) \\
\hline
X-1  & Su1   & PL        & $ 0.36 \pm 0.04 $      & $ 1.83 \pm 0.04 $ & $\cdots$          & $ 5.6 \pm 0.1 $     & 1.16(183) \\
     &       & cutoff-PL & $ 0.27 \pm 0.06 $      & $ 1.5 \pm 0.2 $   & $ 11^{+12}_{-4} $  & $ 5.1^{+0.3}_{-0.2} $ & 1.11(182) \\
     &       & compTT    & $ 0.32^{+0.05}_{-0.04} $ & $ 7.1 $\footnotemark[$\dagger$] & $ 2.4 $\footnotemark[$\dagger$] & $ 5.3^{+0.2}_{-0.3} $ & 1.12(181) \\
     &       & MCD       & $ <0.02 $              & $\cdots$          & $ 1.85 \pm 0.05 $ & $ 4.50 \pm 0.07 $   & 1.70(183) \\
\cline{2-8}
     & Su2   & PL        & $ 0.22 \pm 0.04 $     & $ 1.67 \pm 0.04 $       & $\cdots$          & $ 4.9 \pm 0.1 $     & 1.15(175) \\
     &       & cutoff-PL & $ 0.21^{+0.05}_{-0.06}$ & $ 1.63^{+0.08}_{-0.18} $ & $ >17 $            & $ 4.8^{+0.2}_{-0.1} $ & 1.16(174) \\
     &       & compTT    & $ 0.22 \pm 0.02 $     & $ 6.9 $\footnotemark[$\dagger$] & $ 2.8 $\footnotemark[$\dagger$] & $ 4.8 \pm 0.1 $     & 1.17(173) \\
     &       & MCD       & $ <0.003 $            & $\cdots$                & $ 1.94 \pm 0.05 $ & $ 4.33 \pm 0.07 $   & 2.25(175) \\
\hline
X-2  & Su1   & PL    & $ 3.6 \pm 0.1 $   & $ 2.49 \pm 0.04 $ & $\cdots$          & $ 34 \pm 2 $     & 1.50(396) \\
     &       & MCD   & $ 1.82 \pm 0.08 $ & $\cdots$          & $ 1.78 \pm 0.03 $ & $ 15.0 \pm 0.2 $ & 1.05(396) \\
\cline{2-8}
     & Su2   & PL    & $ 2.2 \pm 0.2 $ & $ 1.71^{+0.07}_{-0.06} $ & $\cdots$            & $ 10.1^{+0.5}_{-0.4} $ & 1.19(175) \\
     &       & MCD   & $ 1.1 \pm 0.1 $ & $\cdots$               & $ 2.7 \pm 0.1 $     & $ 8.9 \pm 0.2 $       & 1.08(175) \\
\hline
\multicolumn{8}{@{}l@{}}{\hbox to 0pt{\parbox{180mm}{\footnotesize
\par\noindent
\footnotemark[$*$] The 0.5--10~keV luminosity for the PL, the cutoff-PL, and the Comptonization models, and the bolometric luminosity for the MCD model.
\par\noindent
\footnotemark[$\dagger$] Only the best-fit value is described for $\tau$ and $k T_{\rm{e}}$, because both parameters are strongly coupled.
}\hss}}
\end{tabular}
\end{center}
\end{table*}

\subsection{Fitting Models}

We used the X-ray spectral fitting package XSPEC version~12.6.0 for the spectral analysis.
We applied two interstellar extinctions:
One is the Galactic extinction represented by the \texttt{wabs} model \citep{mor83}
with a hydrogen column density fixed at the Galactic value toward IC\,342;
$N_{\rm{H}}$~=~3$\times 10^{21}$~cm$^{-2}$ \citep{dic90,sta92}.
The other is a thawed additional extinction for each ULX (as another \texttt{wabs}).

All the spectra are featureless, so we first applied two simple models:
the PL model and the multi-color disk black body model (MCD; \texttt{diskbb} in XSPEC; \cite{mitsu84}),
which represent PL-shaped spectra and convex-shaped spectra, respectively.
These models are often employed for Galactic BHBs and ULXs, including the IC\,342 ULXs (e.g., \cite{kubo01,rob04,mak11}).

If a spectrum is represented by the PL model, we further applied two other models
in order to understand the physical condition of the PL state.
One is the cutoff-PL model, which approximates the Compton emission by constraining two phenomenological parameters:
the photon index and the cutoff (turnover) energy.
The improvement from the PL model is checked by the $F$-test.
We consider that the improvement with the $\ge$99.9\% confidence level is significant.
The other is a Comptonization model (\texttt{compTT} in XSPEC; \cite{tit94}),
which can constrain the electron temperature ($kT_{\rm{e}}$) and the optical depth for electrons ($\tau$) as model parameters.
If the spectrum has no turnover, both parameters are strongly coupled and cannot be constrained.

\subsection{Suzaku Results}

\subsubsection{XIS spectra}

We fitted the FI and BI spectra simultaneously.
The energy range of 1.8--2.0~keV is excluded due to calibration uncertainties.
In addition, for X-2, we did not use data below 1.5~keV to characterize the hard X-ray spectrum,
since a soft excess reported by \citet{feng09} and \citet{mak11} was recognized in the Suzaku data.

\begin{figure*}
\begin{center}
\FigureFile(112mm,150mm){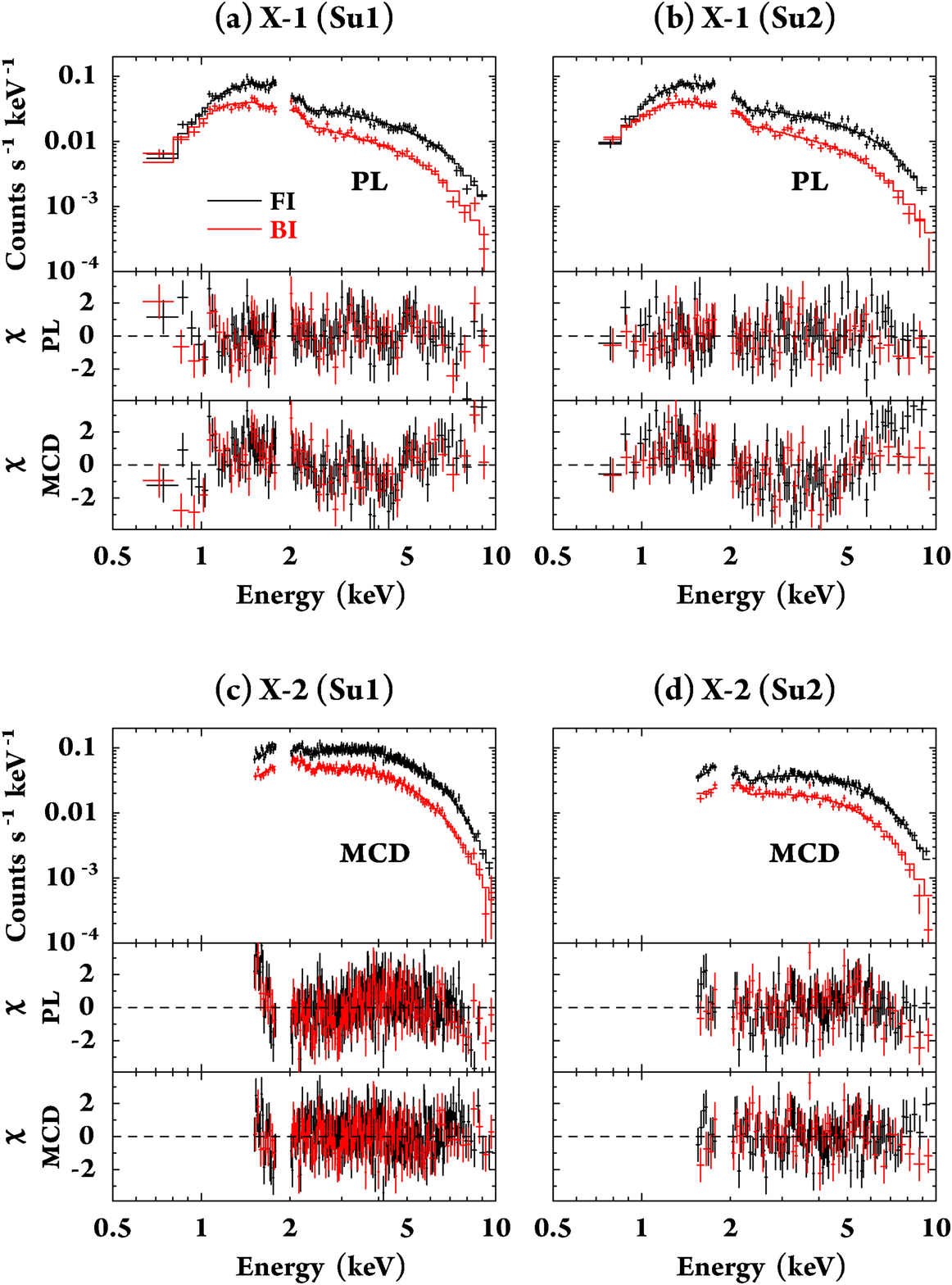}
\end{center}
\caption{Suzaku spectra (pluses) and the best-fit models (solid histogram) of (a--b) X-1 and (c--d) X-2.
Bottom panels show the residuals from the best-fit by a PL and MCD.
\label{fig3}}
\end{figure*}

The Suzaku XIS spectra and the best-fit models are shown in figure~\ref{fig3},
and all the best-fit parameters (90\% confidence level) are summarized in table~\ref{table2}.
The parameters $N^{\prime}_{\rm{H}}$, ${\it{\Gamma}}$, $kT_{\rm{in}}$, and $E_{\rm{cut}}$ indicate
the absorption column density additional to that in our Galaxy,
the photon index, the innermost disk temperature, and the turnover energy, respectively.
The 0.5--10~keV luminosity $L_{\rm{X}}$ for the PL, the cutoff-PL, and the Comptonization models is calculated as $4\pi D^2 f_{\rm{X}}^{\prime}$.
Assuming the disk inclination angle of 60\arcdeg,
the bolometric luminosity $L_{\rm{bol}}$ for the MCD model is calculated as $2\pi D^2 f_{\rm{bol}}^{\prime} (\cos{60\arcdeg})^{-1}$.
Here, $f_{\rm{X}}^{\prime}$ and $f_{\rm{bol}}^{\prime}$ are respectively
the absorption-corrected flux in the 0.5--10~keV and the 0.01--100~keV ranges, and $D$ is the distance to IC\,342 (3.3~Mpc).

\medskip

The X-1 spectra were unaccountable by the MCD model (reduced-$\chi^2$~=~1.70 for Su1 and 2.25 for Su2),
whereas they were successfully fitted by the PL model (reduced-$\chi^2$~=~1.15--1.16).
Based on the derived parameters (table~\ref{table2}), the spectral shape of X-1 slightly changed between the two observations;
in the Su1 observation, the X-ray luminosity and the photon index
were respectively $L_{\rm{X}}$~=~(5.6$\pm$0.1)$\times$10$^{39}$~erg~s$^{-1}$ and ${\it{\Gamma}}$~=~1.83$\pm$0.04,
while in Su2, they were $L_{\rm{X}}$~=~(4.9$\pm$0.1)$\times$10$^{39}$~erg~s$^{-1}$ and ${\it{\Gamma}}$~=~1.67$\pm$0.04.
These parameters are within the typical values (${\it{\Gamma}}$~=~1.6--2.0) measured previously from this ULX;
see, e.g., \citet{kubo01} with ASCA, \citet{feng09} with XMM-Newton, and \citet{rob04} with Chandra.

The cutoff-PL model provided a fit goodness similar to that by the PL model,
but we conclude that both X-1 spectra in Su1 and Su2 do not have a turnover at least below 10~keV,
because the turnover energy ($E_{\rm{cut}}$~=~11$^{+12}_{-4}$~keV for Su1 and $>$17~keV for Su2; table~\ref{table2})
is determined mostly above the XIS fitting range (0.5--10~keV).
In the following ($\S$~4.2.3), we confirm this conclusion by using the PIN data.

The Comptonization model (\texttt{compTT}) fitting is also consistent with the result by the cutoff-PL model (i.e., no turnover),
because $kT_{\rm{e}}$ and $\tau$ were strongly coupled in both spectra, shown in $\S$~5.1.2 for details.
The seed photon temperature was $kT_{\rm{seed}}$~$<$~0.4~keV in both spectra.

\medskip

Unlike X-1, the X-2 spectrum in Su1 cannot be explained by the PL model (reduced-$\chi^2$~=~1.50).
Instead, it has a convex shape represented by the MCD model with a reduced-$\chi^2$ of 1.05 (figure~\ref{fig3}c).
On the other hand, the X-2 spectrum in Su2 was reproduced by both models,
although the MCD model provided a slightly good fitness (figure~\ref{fig3}d; reduced-$\chi^2$~=~1.19 by PL and 1.08 by MCD).
The source spectrum dramatically changed between the two observations; 
the luminosity halved from $L_{\rm{bol}}$~=~(1.50$\pm$0.02)$\times$10$^{40}$ to (8.9$\pm$0.2)$\times$10$^{39}$~erg~s$^{-1}$,
and its disk temperature increased ($kT_{\rm{in}}$~=~1.78$\pm$0.03 to 2.7$\pm$0.1~keV).
Such a spectral behavior is opposite to that of Galactic BHBs in the canonical high-soft or slim disk states.
Its physical interpretation is put aside for the forth-coming paper.

\medskip

\subsubsection{Significance of the source signal with PIN}

For the detailed PIN spectral analysis, we need to consider two background sources:
the non-X-ray background (NXB) and the cosmic X-ray background (CXB).
We adopted NXB spectra provided by the instrument team \citep{fuka09}.
Meanwhile, for CXB, we simulated spectra based on the HEAO-1 model \citep{bol87}.
The count rates of the observed PIN signal, the modeled NXB, the simulated CXB, and the source signal are summarized in table~\ref{table3}.

\begin{figure*}
\begin{center}
\FigureFile(136mm,60mm){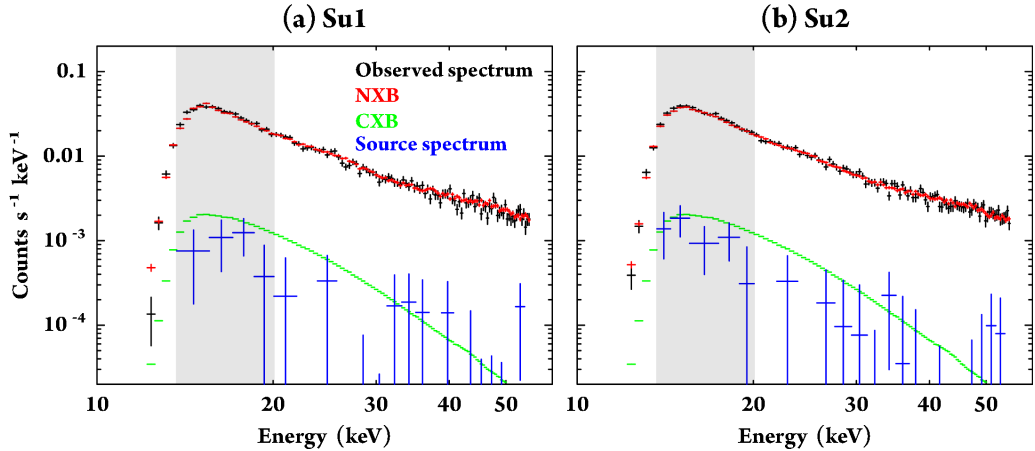}
\end{center}
\caption{Time-averaged signal spectrum (black) and source (blue) spectrum with Suzaku PIN in (a) Su1 and (b) Su2.
The provided NXB and the simulated CXB are represented in red and green, respectively.
The NXB for Su2 is corrected for the rescaling factor ($\S$~4.2.2).
The gray shaded range indicates the notable energy range of 14--20~keV used for the spectral analysis.
\label{fig4}}
\end{figure*}

Normally, the Earth-occultation data with an elevation angle below $-$5\arcdeg\ are utilized to inspect the reproducibility of the NXB model.
However, Suzaku did not undergo the Earth-occultation during our observations.
We, instead, compared the observed PIN signal in the 30--55~keV range with the modeled NXB in the same range,
because such PIN data are considered to be dominated by the NXB events.
Below, we subtracted the CXB contribution from the PIN data to derive the correct rescaling factor.
For Su1, we found no significant difference between the data and the NXB model in this energy range
(the NXB/PIN ratio of 1.5$\pm$2.5\% in 90\% confidence level).
This consistency allows us to adopt for the PIN data in Su1, the NXB model without any correction.
However, the modeled NXB count rate in Su2 is significantly higher than the observed PIN one by 5.5$\pm$2.5\%.
Such a difference is not reasonable from the current uncertainty of the NXB model (typically 1.3\%; \cite{fuka09}).
We therefore decreased the NXB model count rate by 5.5\% for Su2.

\begin{table}
\begin{center}
\caption{Signal and background count rates from Suzaku PIN.\label{table3}}
\small
\begin{tabular}{ccc}
\hline
Name          & \multicolumn{2}{c}{Count rate (s$^{-1}$)\footnotemark[$*$]} \\
\cline{2-3}
              & \multicolumn{2}{c}{Su1} \\
\cline{2-3}
              & 14--20~keV & 30--55~keV \\
\hline
Observed signal   & 0.190$\pm$0.002   & 0.082$\pm$0.001 \\
NXB               & 0.173             & 0.082           \\
CXB               & 0.011             & 0.002           \\
Source net signal & 0.0055$\pm$0.0029 & $\cdots$        \\
\hline
      & \multicolumn{2}{c}{Su2} \\
\cline{2-3}
      & 14--20~keV & 30--55~keV \\
\hline
Observed signal   & 0.187$\pm$0.002                             & 0.081$\pm$0.001                       \\
NXB               & 0.180 (0.170)\footnotemark[$\dagger$]       & 0.084 (0.079)\footnotemark[$\dagger$] \\
CXB               & 0.011                                       & 0.002                                 \\
Source net signal & (0.0068$\pm$0.0028)\footnotemark[$\dagger$] & $\cdots$                              \\
\hline
\multicolumn{3}{@{}l@{}}{\hbox to 0pt{\parbox{85mm}{\footnotesize
\par\noindent
\footnotemark[$*$] The errors are the 1$\sigma$ values.
\par\noindent
\footnotemark[$\dagger$] The parenthetic number indicates the corrected value for the \\
rescaling factor.
}\hss}}
\end{tabular}
\end{center}
\end{table}

Figure~\ref{fig4} shows the ultimately derived PIN spectra.
The source spectrum after subtracting NXB and CXB is represented in blue.
For the spectral analysis, we focus on the 14--20~keV band.
The net count rates (after the NXB and CXB were subtracted) of Su1 and Su2 are calculated as
(5.5$\pm$2.9)$\times$$10^{-3}$~counts~s$^{-1}$ and (6.8$\pm$2.8)$\times$$10^{-3}$~counts~s$^{-1}$, respectively (table~\ref{table3}).
These net count rates are de-convolved to the X-ray flux in the same band
as (1.2$\pm$0.6)$\times$$10^{-12}$~erg~cm$^{-2}$~s$^{-1}$ and (1.4$\pm$0.6)$\times$$10^{-12}$~erg~cm$^{-2}$~s$^{-1}$ for Su1 and Su2, respectively.
Here, the errors of the PIN count rate and the de-convolved PIN flux are the 1$\sigma$ values,
in which the statistical and systematic uncertainties of the NXB were taken into account.

\medskip

Next, we confirm that the majority of the detected (net) PIN fluxes are from the two ULXs by inspecting two X-ray catalogs:
First, we used the Swift BAT 54~months catalog \citep{cus10}, which lists no source in our region,
and calculated the probability that sources (other than X-1 and X-2) possibly contributing to the PIN flux are located within the PIN view.
The detailed procedure is performed in the following two steps:
(i) The flux limit of the catalog is 1$\times$10$^{-11}$~erg~cm$^{-2}$~s$^{-1}$ in 15--150~keV.
In order to compare with the PIN flux limit ($\sim$5$\times$10$^{-13}$~erg~cm$^{-2}$~s$^{-1}$ in 14--20~keV),
we calculated the 14--20~keV flux limit of the catalog by assuming a PL shape with ${\it{\Gamma}}$~=~2.0.
Its value is 1.5$\times$10$^{-12}$~erg~cm$^{-2}$~s$^{-1}$, which is greater than that of PIN.
The fact indicates that sources (including the two ULXs) with a 14--20~keV flux below the catalog flux limit contribute the detected PIN flux.
(ii) By extrapolating the logN--logS relation of the catalog (without Galactic sources at $|l|$~$<$~10\arcdeg),
we derived that the expected number of sources with a 14--20~keV flux between the flux limit of the catalog and that of PIN is about 2,000 in all sky.
From this number and the area ratio of the PIN view and all sky,
we conclude the probability that such three or more sources are simultaneously located within the PIN view to be at most 0.1\%.

Second, we used the Chandra catalog \citep{liu11}
and examined the total 14--20~keV flux of 56 X-ray point-like sources (other than X-1 and X-2) within the IC\,342 region.
Each source shows the 0.3--8.0~keV flux of $10^{-12}$ to $10^{-15}$~erg~cm$^{-2}$~s$^{-1}$.
When a PL shape with ${\it{\Gamma}}$~=~2.0 is assumed, the 14--20~keV flux of each source is calculated as 10$^{-13}$ to 10$^{-16}$~erg~cm$^{-2}$~s$^{-1}$.
Consequently, its total flux is summed up $\sim$2.9$\times$10$^{-13}$~erg~cm$^{-2}$~s$^{-1}$, which corresponds to at most 20\% of the detected PIN flux.

\begin{figure*}
\begin{center}
\FigureFile(160mm,48mm){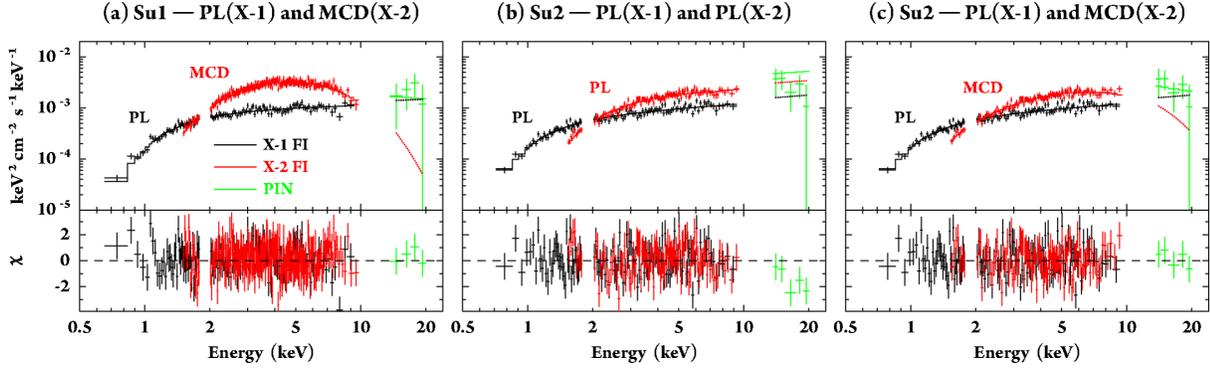}
\end{center}
\caption{XIS plus PIN joint unfolded spectra (pluses).
Black and red are respectively XIS spectra of X-1 and X-2, while green is PIN one.
(a) The spectral models (histograms) are PL for X-1 and MCD for X-2 in Su1,
(b) PL for X-1 and PL for X-2 in Su2, and (c) PL for X-1 and MCD for X-2 in Su2.
\label{fig5}}
\end{figure*}

\medskip

\subsubsection{XIS + PIN spectra}

In the XIS analysis ($\S$~4.2.1), all the XIS spectra were successfully fitted by a phenomenological model (PL, MCD, or cutoff-PL).
Now, we investigate spectral shapes above 10~keV of the ULXs
by extrapolating the successful XIS models (with the best-fit central parameters in table~\ref{table2}) to the PIN band.
We note that X-1 and X-2 are treated simultaneously, since the PIN spectrum contains the signals from both ULXs.
The PIN effective areas to each ULX are much the same\footnote{
See ``The Suzaku Technical Description'' available at http://www.astro.isas.jaxa.jp/suzaku/doc/suzaku\_td\_ao7.pdf for details.}.
We took into account the known calibration uncertainty between XIS and PIN normalizations\footnote{
See http://www.astro.isas.ac.jp/suzaku/doc/suzakumemo/suzakumemo-2008-06.pdf for details.},
namely the PIN normalization is larger than the XIS one by a factor of 1.164.

Table~\ref{table4} shows the spectral model components for X-1 and X-2.
We check the validity of the extrapolation by comparing the modeled flux with the de-convolved PIN flux in 14--20~keV.
For Su1, we examined the PL(X-1)$+$MCD(X-2) combination and the cutoff-PL$+$MCD one ($E_{\rm{cut}}$~=~11~keV for X-1 is assumed).
Clearly, the PIN spectrum was better reproduced by the former combination (figure~\ref{fig5}a) than the latter.
This is because the modeled flux of the former (0.91$\times$10$^{-12}$~erg~cm$^{-2}$~s$^{-1}$) is comparable to
the de-convolved PIN flux of (1.2$\pm$0.6)$\times$10$^{-12}$~erg~cm$^{-2}$~s$^{-1}$,
while the flux of the latter (0.52$\times$10$^{-12}$~erg~cm$^{-2}$~s$^{-1}$) is smaller than the lower limit (1$\sigma$) of the de-convolved flux.
Consequently, the X-1 spectrum in Su1 with no turnover (below 20~keV) was confirmed as mentioned in $\S$~4.2.1.
For Su2, we examined the PL and MCD models to X-2 since its XIS spectrum was reproduced by both models ($\S$~4.2.1),
while only the PL model was considered to X-1.  
Figure~\ref{fig6}~(b) clearly indicates that
the PL$+$PL combination with a flux of 2.9$\times$10$^{-12}$~erg~cm$^{-2}$~s$^{-1}$ exceeds
the PIN spectrum with (1.4$\pm$0.7)$\times$10$^{-12}$~erg~cm$^{-2}$~s$^{-1}$.
By replacing the PL component with the MCD one for X-2,
the model flux of 1.3$\times$10$^{-12}$~erg~cm$^{-2}$~s$^{-1}$ became consistent with the observed one, as displayed in figure~\ref{fig6}~(c).

Based on the present joint analysis, we consider the spectral properties as follows:
(i) In both observations, the X-1 spectra have a PL tail extending up to at least 20~keV.
(ii) Both of the X-2 spectra cannot be explained by a single PL model.
They instead have a convex shape.

\subsection{XMM-Newton, Chandra, and Swift Results}

We analyzed the spectra obtained with the three satellites using the method same as for the Suzaku XIS ($\S$~4.2.1).
For the XMM-Newton EPIC data, we fitted the MOS and pn spectra simultaneously.
The MOS spectra were multiplied by a constant factor of 1.0--1.1,
because \citet{mat09} reported that the MOS flux is higher than pn by 7--9\% ($<$4.5~keV) and 10--13\% ($\gtrsim$4.5~keV).
The best-fit parameters are summarized in table~\ref{table5}.

\medskip

Same as the Suzaku results,
most X-1 spectra (XM1, XM3, Ch1, Ch2, Sw1, and Sw2$+$3) were better reproduced by the PL model.
They were also reproduced by the cutoff-PL model, but we did not derive a significant improvement.
Their X-ray luminosities (except for Sw1) are comparable to that in the Suzaku observations ($\sim$5$\times 10^{39}$~erg~s$^{-1}$).
In the Comptonization model fitting for these spectra,
$kT_{\rm{e}}$ and $\tau$ were coupled, and the upper limit of the seed photon temperature was determined at 0.5~keV.

\begin{table}
\begin{center}
\caption{Modeled and de-convolved fluxes in 14--20~keV.\label{table4}}
\footnotesize
\begin{tabular}{ccccccc}
\hline
Data  & \multicolumn{2}{c}{Model} & & \multicolumn{2}{c}{Flux in 14--20~keV\footnotemark[$*$]} & Notes \\
\cline{2-3} \cline{5-6}
label & X-1 & X-2                 & & Model & Net\footnotemark[$\dagger$] & \\
\hline
Su1 & PL        & MCD & & 0.91 & 1.2$\pm$0.6 & figure~\ref{fig5}~(a) \\
Su1 & cutoff-PL & MCD & & 0.52 & 1.2$\pm$0.6 & --- \\
Su2 & PL        & PL  & & 2.9  & 1.4$\pm$0.6 & figure~\ref{fig5}~(b) \\
Su2 & PL        & MCD & & 1.3  & 1.4$\pm$0.6 & figure~\ref{fig5}~(c) \\
\hline
\multicolumn{7}{@{}l@{}}{\hbox to 0pt{\parbox{85mm}{\footnotesize
\par\noindent
\footnotemark[$*$] The PIN band flux in unit of $10^{-12}\;\rm{erg \; cm^{-2} \; s^{-1}}$.
\par\noindent
\footnotemark[$\dagger$] The de-convolved PIN flux from the net count rate obtained by subtracting the NXB and CXB from the observed PIN count rate.
The errors are the 1$\sigma$ values.
}\hss}}
\end{tabular}
\end{center}
\end{table}

\begin{figure}
\begin{center}
\FigureFile(68mm,116mm){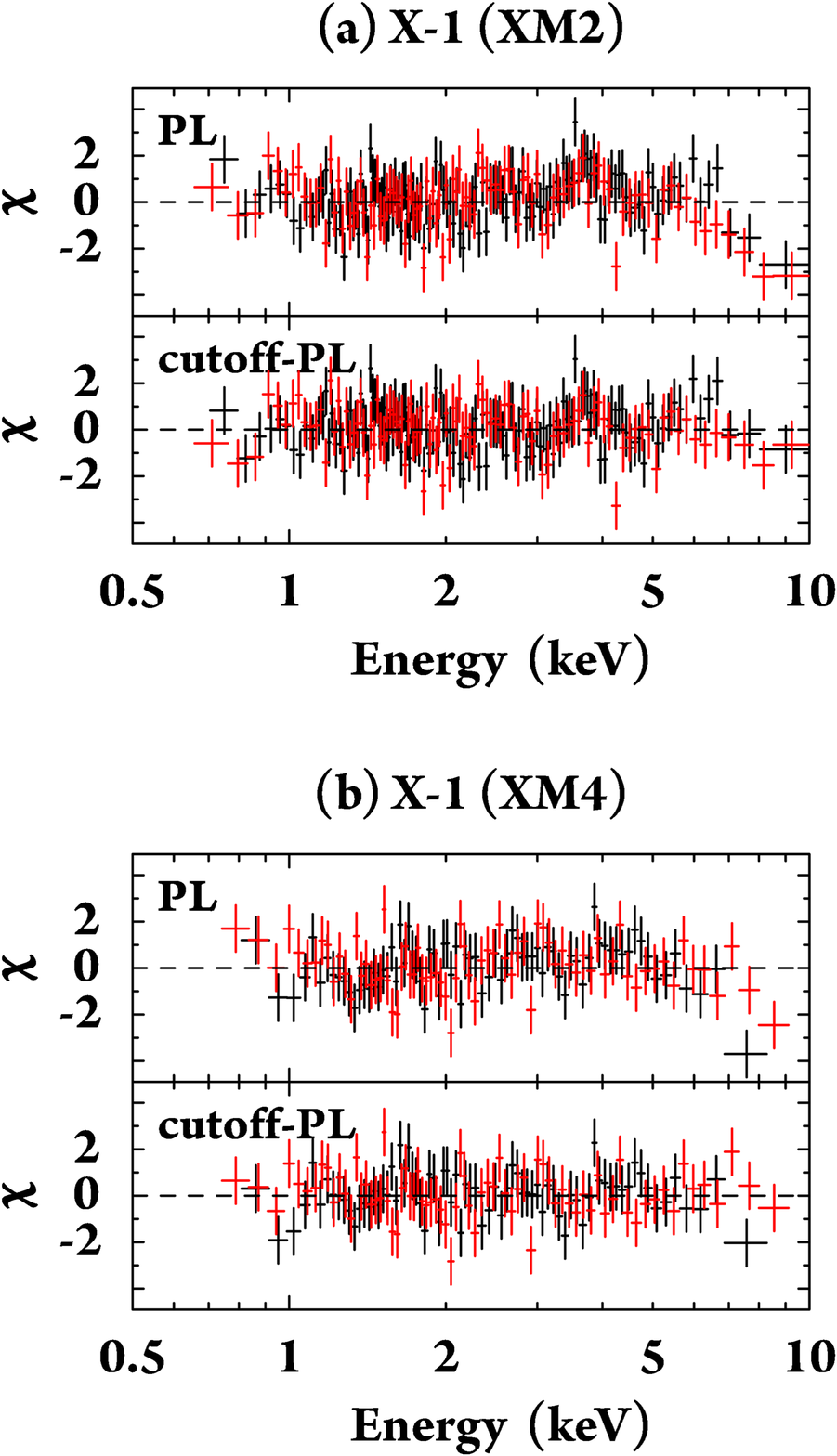}
\end{center}
\caption{Residuals from the best-fit by the PL model and the cutoff-PL model in (a) XM2 and (b) XM4.
Black and red represent MOS and pn, respectively.
\label{fig6}}
\end{figure}

On the other hand, for the two spectra in XM2 and XM4,
the PL model yielded negative residuals particularly at $\gtrsim$6~keV, indicating the existence of a turnover (figure~\ref{fig6}).
In fact, these residuals were significantly improved by applying the cutoff-PL model (figure~\ref{fig6} and table~\ref{table5}).
The reduced-$\chi^2$(d.o.f.) is from 1.18(233) by PL to 0.97(232) by cutoff-PL in XM2, and from 1.12(147) by PL to 0.93(146) by cutoff-PL in XM4,
indicating the $F$-test significance of $>$99.9\%.
The derived turnover energies are $E_{\rm{cut}}$~=~$6.1^{+1.9}_{-1.2}$~keV in XM2 and $5.1^{+2.3}_{-1.2}$~keV in XM4.
Their X-ray luminosities ($\sim$1$\times 10^{40}$~erg~s$^{-1}$) are higher than
the other (most) observations with no turnover blow 10~keV (or 20~keV with Suzaku) by a factor of 2 to 3.
These results suggest that the spectral turnover (below 10~keV) is prominent only in the higher luminosity phase.
In addition, for such spectra (XM2 and XM4), the Compton parameters were determined as $kT_{\rm{e}}$~$\sim$~1.8~keV and $\tau$~$\sim$~8.5,
although the upper limit of the seed photon temperature was 0.5~keV like the other observations.

\begin{table*}
\begin{center}
\caption{Best-fit parameters for the XMM-Newton, Chandra, and Swift spectra.\label{table5}}
\small
\begin{tabular}{cccccccc}
\hline
Name & Data  & Model & $N^{\prime}_{\rm{H}}$      & ${\it{\Gamma}}$ or $\tau$ & $E_{\rm{cut}}$, $k T_{\rm{e}}$, or $k T_{\rm{in}}$ & Luminosity\footnotemark[$*$]     & Red--$\chi^2$ \\
     & label &       & $(10^{22}\;\rm{cm}^{-2})$ &                           & (keV)                                           & $(10^{39}\;\rm{erg \; s^{-1}})$ & (d.o.f.) \\
\hline
X-1  & XM1   & PL        & $ 0.24^{+0.07}_{-0.06} $ & $ 1.63 \pm 0.09 $   & $\cdots$          & $ 4.0^{+0.2}_{-0.1} $ & 1.00(90) \\
     &       & cutoff-PL & $ 0.22^{+0.07}_{-0.11} $ & $ 1.6^{+0.1}_{-0.4} $ & $ >7.8 $          & $ 4.0^{+0.2}_{-0.4} $ & 1.01(89) \\
     &       & compTT    & $ <0.26 $              & $ 8.5 $\footnotemark[$\dagger$] & $ 2.3 $\footnotemark[$\dagger$] & $ 3.9 \pm 0.3 $     & 0.97(88) \\
     &       & MCD       & $ <0.02 $              & $\cdots$             & $ 1.9 \pm 0.1 $   & $ 3.9 \pm 0.2 $     & 1.29(90) \\
\cline{2-8}
     & XM2   & PL        & $ 0.53^{+0.03}_{-0.02} $ & $ 2.02 \pm 0.03 $   & $\cdots$            & $ 11.1 \pm 0.2 $     & 1.18(233) \\
     &       & cutoff-PL & $ 0.39 \pm 0.04 $      & $ 1.4^{+0.1}_{-0.2} $ & $ 6.1^{+1.9}_{-1.2} $ & $ 9.4 \pm 0.4 $    & 0.97(232) \\
     &       & compTT    & $ 0.27^{+0.20}_{-0.08} $ & $ 8.5^{+0.6}_{-1.4} $ & $ 1.8 \pm 0.1 $     & $ 8.2^{+1.8}_{-0.4} $ & 0.95(231) \\
     &       & MCD       & $ 0.15 \pm 0.02 $      & $\cdots$             & $ 1.54 \pm 0.03 $   & $ 9.2 \pm 0.1 $     & 1.72(233) \\
\cline{2-8}
     & XM3   & PL        & $ 0.30 \pm 0.03 $ & $ 1.82 \pm 0.05 $      & $\cdots$          & $ 4.8 \pm 0.1 $ & 1.06(123) \\
     &       & cutoff-PL & $ 0.30 \pm 0.03 $ & $ 1.81^{+0.05}_{-0.08} $ & $ >0.01 $         & $ 4.8 \pm 0.1 $ & 1.07(122) \\
     &       & compTT    & $ 0.29 \pm 0.03 $ & $ 5.0 $\footnotemark[$\dagger$] & $ 4.3 $\footnotemark[$\dagger$] & $ 5.2 \pm 0.1 $ & 1.08(121) \\
     &       & MCD       & $ <0.01 $         & $\cdots$               & $ 1.78 \pm 0.05 $ & $ 4.2 \pm 0.1 $ & 2.69(123) \\
\cline{2-8}
     & XM4   & PL        & $ 0.55 \pm 0.04 $ & $ 1.91 \pm 0.05 $ & $\cdots$            & $ 14.2 \pm 0.4 $      & 1.12(147) \\
     &       & cutoff-PL & $ 0.37 \pm 0.07 $ & $ 1.2 \pm 0.2 $   & $ 5.1^{+2.3}_{-1.2} $ & $ 11.8^{+0.7}_{-0.6} $ & 0.93(146) \\
     &       & compTT    & $ 0.46 \pm 0.05 $ & $ 8.4 \pm 0.9 $   & $ 1.8^{+0.3}_{-0.2} $ & $ 12.7^{+0.4}_{-0.3} $  & 0.93(145) \\
     &       & MCD       & $ 0.17 \pm 0.03 $ & $\cdots$          & $ 1.71 \pm 0.07 $   & $ 12.1 \pm 0.2 $      & 1.22(147) \\
\cline{2-8}
     & Ch1   & PL        & $ 0.3 \pm 0.1 $ & $ 1.8 \pm 0.2 $      & $\cdots$        & $ 4.7 \pm 0.3 $     & 0.98(25) \\
     &       & MCD       & $ <0.05 $       & $\cdots$             & $ 1.7 \pm 0.2 $ & $ 3.7 \pm 0.2 $     & 1.54(25) \\
     &       & cutoff-PL & $ 0.3 \pm 0.1 $ & $ 1.8^{+0.2}_{-0.3} $ & $ >0.01 $       & $ 4.6^{+0.3}_{-0.7} $ & 1.02(24) \\
     &       & compTT    & $ 0.3 \pm 0.1 $ & $ 4.5 $\footnotemark[$\dagger$] & $ 5.4 $\footnotemark[$\dagger$] & $ 4.6 \pm 0.3 $     & 1.07(23) \\
\cline{2-8}
     & Ch2   & PL        & $ 0.3 \pm 0.1 $      & $ 1.7^{+0.2}_{-0.1} $ & $\cdots$        & $ 5.0 \pm 0.3 $     & 0.57(27) \\
     &       & cutoff-PL & $ 0.3^{+0.1}_{-0.2} $ & $ 1.7^{+0.2}_{-0.7} $ & $ >0.01 $       & $ 5.0^{+0.3}_{-0.8} $ & 0.59(26) \\
     &       & compTT    & $ 0.3 \pm 0.1 $      & $ 4.2 $\footnotemark[$\dagger$] & $ 6.4 $\footnotemark[$\dagger$] & $ 5.0 \pm 0.3 $  & 0.62(25) \\
     &       & MCD       & $ <0.08 $            & $\cdots$            & $ 1.8 \pm 0.2 $ & $ 4.0 \pm 0.2 $     & 0.98(27) \\
\cline{2-8}
     & Sw1   & PL        & $ 0.5^{+0.3}_{-0.2} $ & $ 1.8 \pm 0.3 $  & $\cdots$            & $ 13^{+2}_{-1} $ & 0.69(29) \\
     &       & cutoff-PL & $ <0.7 $             & $ <1.9 $        & $ 0.001 $--$ 0.002 $ & $ 9^{+4}_{-2} $  & 0.64(28) \\
     &       & compTT    & $ <0.7 $             & $ 11.4 $\footnotemark[$\dagger$] & $ 1.4 $\footnotemark[$\dagger$] & $ 10^{+3}_{-2} $ & 0.66(27) \\
     &       & MCD       & $ <0.3 $             & $\cdots$        & $ 1.8^{+0.4}_{-0.3} $ & $ 10 \pm 1 $    & 0.61(29) \\
\cline{2-8}
     & Sw2+3 & PL        & $ 0.5^{+0.4}_{-0.3} $ & $ 1.8 \pm 0.4 $ & $\cdots$             & $ 4.9^{+1.1}_{-0.7} $ & 0.72(12) \\
     &       & cutoff-PL & $ <0.9 $             & $ <2.2 $        & $ >2.1 $             & $ 5^{+1}_{-2} $      & 0.78(11) \\
     &       & compTT    & $ <0.8 $             & $ 4.9 $\footnotemark[$\dagger$] & $ 5.3 $\footnotemark[$\dagger$] & $ 4^{+5}_{-1} $      & 0.83(10) \\
     &       & MCD       & $ <0.3 $             & $\cdots$        & $ 1.8^{+0.5}_{-0.4} $ & $ 3.7^{+0.5}_{-0.4} $ & 0.91(12) \\
\hline
\multicolumn{8}{@{}l@{}}{\hbox to 0pt{\parbox{180mm}{\footnotesize
\par\noindent
\footnotemark[$*$] Same as table~\ref{table2}.
\par\noindent
\footnotemark[$\dagger$] Same as table~\ref{table2}.
}\hss}}
\end{tabular}
\end{center}
\end{table*}

\section{Discussion}

In the present study, we examine the X-ray emission properties of the ULX IC\,342 X-1,
aiming at understanding the nature of accretion flow in ULXs during the PL state.
In contrast with the convex-shaped (thermal) spectra, from which we can easily obtain concrete information,
such as the mass accretion rate and the BH spin (as long as the blackbody approximation is adopted),
it is not so easy to do so for the PL spectra.
Nevertheless, we retrieved some information from the PL slope, the existence (or absence) of a turnover,
a turnover energy (if a turnover exists), its timing properties, and so on.
We will pay special attention to the turnover energy, since it provides a key to understanding the nature of the PL state.

\subsection{Summary of the Observations of IC\,342 X-1}

\subsubsection{Luminosities and spectral shapes}

\begin{figure*}
\begin{center}
\FigureFile(136mm,57mm){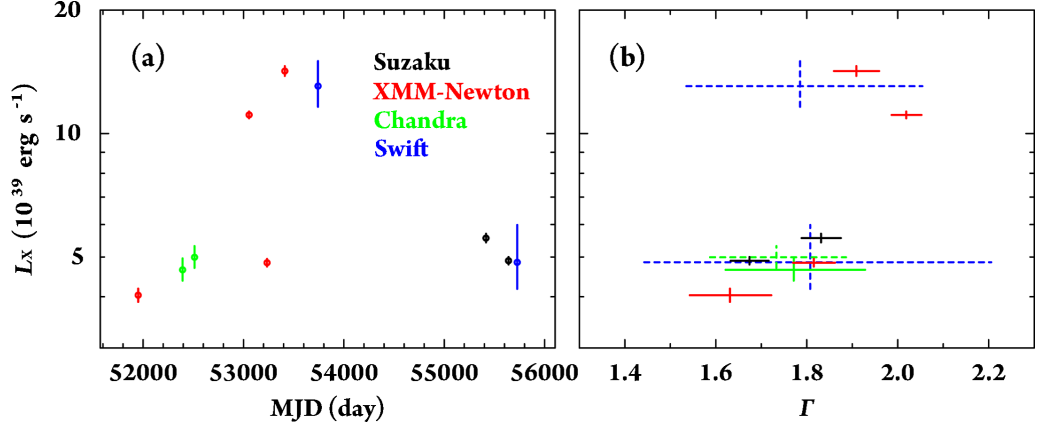}
\end{center}
\caption{(a) Long-term X-ray variability,
and (b) $L_{\rm{X}}$ (X-ray luminosity) versus ${\it{\Gamma}}$ (photon index) diagram of IC\,342 X-1.
The parameters are derived by spectral fitting with the PL model.
The data taken by the different satellites are indicated by the different colors.
\label{fig7}}
\end{figure*}

We plot the X-ray luminosity (in the energy range of 0.5--10~keV) against the observed date and the photon index in figure~\ref{fig7}.
We can see two groups of data occupying separate regions, as was reported previously \citep{feng09,mak11}, in these plots:
the low luminosity group with (4--6)$\times 10^{39}$~erg~s$^{-1}$ (Su1, Su2, XM1, XM3, Ch1, Ch2, and Sw2$+$3)
and the high luminosity group with (1.1--1.4)$\times 10^{40}$~erg~s$^{-1}$ (XM2, XM4, and Sw1).
Since both groups have been called the ``PL'' state previously,
we expediently hereafter call them the low luminosity PL state and the high luminosity PL state respectively,
although the spectra in the high luminosity PL state have a turnover (see below).
No significant difference was seen in the photon indices between the two PL states.
The unpublished data (Su1, Su2, Sw1, and Sw2$+$3) also follow these trends.
Although the strict duty cycle cannot be estimated with this sparse data sampling,
it seems that the low luminosity PL state appears more frequently than the high luminosity PL state.
Its appearing fraction is low:high~=~7:3 based on the number of the present samples in each state.

\begin{table*}
\begin{center}
\caption{Basic properties of the two PL states of IC\,342 X-1.\label{table6}}
\small
\begin{tabular}{lll}
\hline
\multicolumn{1}{c}{Properties} & \multicolumn{2}{c}{Luminosity State} \\
\cline{2-3}
                               & \multicolumn{1}{c}{Low PL} & \multicolumn{1}{c}{High PL} \\
\hline
Data set                             & Su1, Su2, XM1, XM3, Ch1, Ch2, Sw2$+$3 & XM2, XM4, Sw1 \\
Appearing fraction (present samples) & 70\%                                  & 30\% \\
Range of $L_{\rm{X}}$ (erg~s$^{-1}$)   & (4--6)$\times$$10^{39}$               & (1.1--1.4)$\times$$10^{40}$ \\
Turnover energy                      & Unconstraint ($>$20~keV)              & $\sim$6~keV \\
\hline
\end{tabular}
\end{center}
\end{table*}

\begin{figure}
\begin{center}
\FigureFile(72mm,51mm){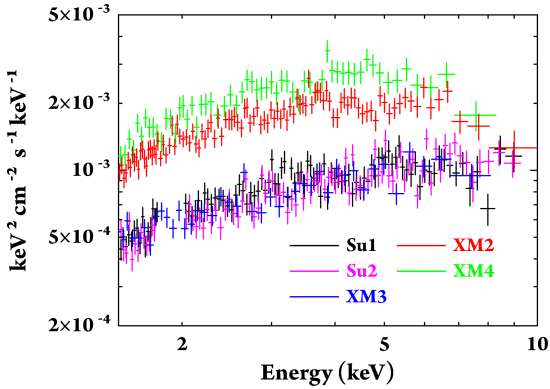}
\end{center}
\caption{Unfolded spectra of IC\,342 X-1 in Su1, Su2, XM2, XM3, and XM4 shown with different colors.
\label{fig8}}
\end{figure}

Besides the luminosity differences, X-1 also showed somewhat different spectral features in hard X-rays between these two states.
Figure~\ref{fig8} displays the unfolded $\nu$$F_{\rm{\nu}}$ spectra in the unit of keV$^2$~cm$^{-2}$~s$^{-1}$~keV$^{-1}$
in five observations (Su1--2 and XM2--4 with good statistics).
Evidently, the spectral shapes differ at $\gtrsim$6~keV.
During the high luminosity PL state (XM2 and XM4), the spectrum has a turnover at around 6~keV.
We can confirm this property by comparing the spectral fittings to the data with the PL model and with the cutoff-PL model.
The spectra in the high luminosity PL state are significantly improved, when we apply the cutoff-PL model with the $F$-test significance at $>$99.9\% ($\S$~4.3).
Whereas, during the low luminosity PL state (Su1, Su2, and XM3), the spectrum has no turnover below 10~keV.
In fact, the spectra are not improved by the cutoff-PL model ($\S$~4.2.1 and $\S$~4.3).
We estimate that a PL tail extends up to at least 20~keV,
which we firstly confirmed by our Suzaku PIN detector observation ($\S$~4.2.3).
We expect similar hard X-ray emissions from this source in the other observations during the low luminosity PL state,
since the derived values ($L_{\rm{X}}$ and ${\it{\Gamma}}$ in figure~\ref{fig7}b) are similar,
although they were too faint to detect with satellites other than Suzaku.
We thus conclude that there exists a certain trend in the spectral changes with an increase in $L_{\rm{X}}$.
Table~\ref{table6} summarizes the basic properties of the two PL states.

\medskip

\subsubsection{Spectral fitting with the Comptonization model}

The PL-like spectra of Galactic BHBs during the VHS are usually explained in terms of the Comptonization model \citep{done07}.
From an overall spectral similarity,
we may assume that the PL-like spectra in both PL states are generated by the inverse Compton scattering of soft photons.
However, cautions should be taken here,
since there exist some differences in the observed properties between the PL states of IC\,342 X-1 and the VHS of Galactic BHBs;
that is, (i) the photon indices of the PL states (${\it{\Gamma}}$~=~1.6--2.0) are smaller than that of the VHS ($>$2.4; \cite{mcc06}).
(ii) The spectral turnover at a lower energy, around 6~keV, found in the high luminosity PL state is not observed during the VHS.
(iii) The significant variability observed in the VHS is not confirmed in the PL states.

\begin{figure*}
\begin{center}
\FigureFile(136mm,60mm){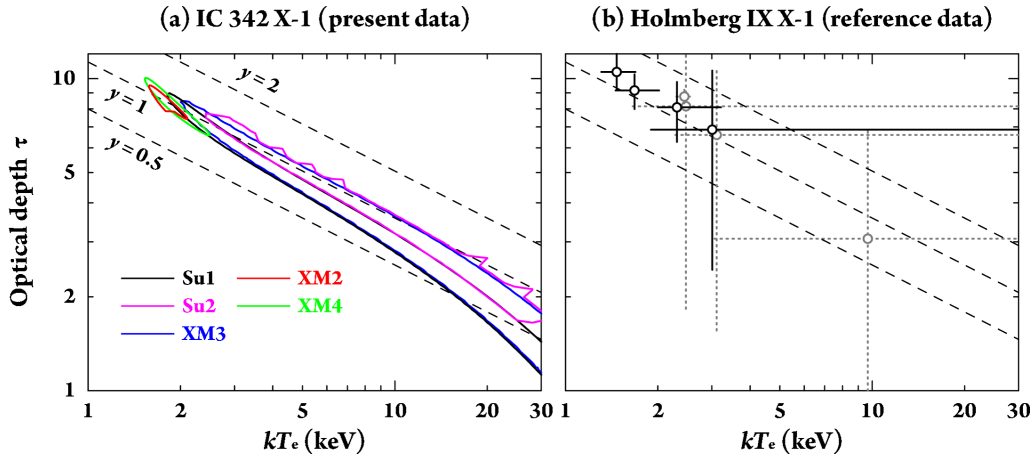}
\end{center}
\caption{(a) Significance contours of the best-fit physical parameters ($\tau$ versus $kT_{\rm{e}}$) by the Comptonization model for IC\,342 X-1.
The results in different observations are shown with different colors: 
Su1 (black), Su2 (magenta), XM2 (red), XM3 (blue), and XM4 (green).
The confidence level is shown for the 3$\sigma$ range.
The dashed lines represent the constant Compton $y$-parameter of 0.5, 1, and 2.
(b) $\tau$ against $kT_{\rm{e}}$ of Holmberg\,XI X-1 by \citet{kiki10a}.
The data in black (solid line) show a generally trend to increase in the optical depth as the X-ray luminosity increases,
while the data in gray (dotted line) do not.
The bolometric luminosities in black from top to bottom are 11.1, 9.41, 6.92, and 6.63$\times 10^{39}$~erg~s$^{-1}$,
and those in gray from top to bottom are 5.92, 17.3, 9.27, and 11.1$\times 10^{39}$~erg~s$^{-1}$.
\label{fig09}}
\end{figure*}

In order to understand the physical condition of the two PL states,
we show, in figure~\ref{fig09}~(a), the best-fit confidence contours between $\tau$ and $kT_{\rm{e}}$ derived from the Comptonization model,
for the five observations (Su1--2 and XM2--4).
Some specific values of Compton $y$-parameters ($y$~=~0.5, 1, and 2) are also indicated in this figure with the dashed lines.

There are two important facts seen in figure~\ref{fig09}~(a):
First, the two states occupy different regions in this $\tau$--$kT_{\rm{e}}$ plot:
$kT_{\rm{e}}$~$\sim$~1.8~keV and $\tau$~$\sim$~8.5 in the high luminosity PL state,
while $kT_{\rm{e}}$ ($>$2~keV) and $\tau$ ($<$8) are unconstrained in the low luminosity PL state.
Second, despite different values of $kT_{\rm{e}}$ and $\tau$,
both states show the region with similar $y$-values; i.e., between $y$~=~0.5 and 2.
This is a noteworthy feature and its reason should be considered carefully,
since the $y$-parameter is closely connected to the physical condition of a Comptonizing corona (see the next subsection).

Note that PL-like spectra are produced by unsaturated inverse Compton scattering
in the energy range of $kT_{\rm{seed}}$~$\ll$~$h\nu$~$\ll$~$kT_{\rm{e}}$
on the condition that the seed photon temperature ($kT_{\rm{seed}}$) is much less than the electron temperature of a Comptonizing corona.
This is the case that we encounter here,
since from the spectral fitting we estimate $kT_{\rm{seed}}$~$\lesssim$~0.5~keV, while $kT_{\rm{e}}$~$>$~1~keV in the high luminosity PL state.
A similar argument applies to the low luminosity PL state, as well.

In the following subsection, we interpret the PL states based on these spectral properties.

\subsection{Physical Interpretations of the Two PL States}

\subsubsection{The high luminosity PL state --- possible scenario invoking a Comptonizing corona}

First, we consider the high luminosity PL state, which is characterized by a turnover below 10~keV.
\citet{kubo02} firstly reported the existence of a turnover below 10~keV in the PL-shaped spectrum of a ULX (IC\,342 X-1) with ASCA.
\citet{miyawa09} also reported a turnover in the PL-shaped spectrum of another ULX (M\,82 X-1) with Suzaku.
They interpreted that such ULXs are in a state corresponding to the ``normal'' VHS of Galactic BHBs.
We, however, already pointed out some distinctions between the VHS of Galactic BHBs and the high luminosity PL state (see $\S$~5.1.2).

The PL-shaped spectra with a turnover below 10~keV were reported in the high quality spectra of several ULXs analyzed by \citet{gla09}.
They found that such spectra can be explained by the emission from a low-temperature optically-thick corona with $kT_{\rm{e}}$~$<$~2.5~keV and $\tau$~$>$~6.
They proposed to call it the ultraluminous state (see \cite{sor07}),
claiming that the low-temperature optically-thick corona is formed by supercritical accretion flows.
Their idea has been confirmed with global radiation-hydrodynamic simulations by \citet{kawa12}.
\citet{kiki10a} also reported the results of the PL-shaped spectra with a turnover,
which they have seen from a long-term monitoring observation of the ULX Holmberg\,IX X-1 (one of the ULXs analyzed by \cite{gla09})
taken with XMM-Newton and Swift.
They discovered a general trend of decreasing electron temperatures and increasing optical depths as the X-ray luminosity increases,
especially at $\tau$~$>$~6 (the black solid lines in figure~\ref{fig09}b).
Notably, its $y$-parameter is $\sim$1.
This $\tau$--$kT_{\rm{e}}$ relation (including $y$-parameter) is similar to that of IC\,342 X-1.

From these similarities, we suggest that IC\,342 X-1 is in a similar physical condition to that of
some ULXs reported by \citet{gla09} and \citet{kiki10a}.
In other words, IC\,342 X-1 (at least during the high luminosity PL state) is likely to undergo supercritical accretion flows,
if the interpretation by \citet{gla09} and \citet{kawa12} is correct.

However, \citet{kiki10a} also reported somewhat different behavior.
First, Holmberg\,IX X-1 showed rather continuous spectral variations,
while IC\,342 X-1 seems to stay in two separate states, although the data sampling is rather sparse.
Second, spectral changes of Holmberg\,IX X-1 are not always uniquely determined by its X-ray luminosity
(see the gray dotted lines in figure~\ref{fig09}b).
It is interesting to point that Galactic BHBs exhibit hysteretic spectral transitions;
their spectral states differ in the rising and decaying phases (see, e.g., \cite{miyamo93}; \cite{fend04}).
We may expect similar hysteretic behavior, and if so, distinct spectral states appear even at similar luminosities.
Certainly, we need more continuous observations to see if these differences really exist.

We briefly comment on the case of the Galactic BHB GRS\,1915$+$105,
in which the mass of the BH is known to be 14$\pm$4~$M_{\odot}$ \citep{gre01}.
At near the Eddington luminosity (apparent $L$~$\sim$~0.7$L_{\rm{Edd}}$) under the supercritical accretion situation,
the source showed a unique state with a low electron temperature ($kT_{\rm{e}}$~$\sim$~3~keV)
and optically-thick ($\tau$~$\sim$~9) Comptonizing corona \citep{kiki10b}.
Such features are quite reminiscent of those of the PL state with a turnover of ULXs,
and this similarity may suggest similar physical conditions;
i.e., the supercritical accretion may also be occurring in the ULXs.
We need to point an annoying fact, however; that is the seed photon temperatures are distinct:
GRS\,1915$+$105 exhibits a disk-dominated spectrum with $kT_{\rm{seed}}$~$\sim$~2~keV,
while our data of IC\,342 X-1 show much lower seed photon temperatures, below 0.5~keV.
The origin of this discrepancy is left as an open question.

\medskip

\subsubsection{A possible theoretical interpretation of the high luminosity PL state}

Here, we try to understand a physical cause of yielding $y$~$\sim$~1 observed in the high luminosity PL state based on the framework of the disk-corona model.
Let us consider the energy balance of the corona and the accretion disk.
The former can be described by $Q^{+}_{\rm{corona}}$~=~$Q^{-}_{\rm{corona}}$,
where $Q^{+}_{\rm{corona}}$ and $Q^{-}_{\rm{corona}}$ represent the heating and the cooling rates of the corona, respectively.
The latter is approximated by $Q^{-}_{\rm{corona}}$~$\sim$~$(y/2) c U_{\rm{seed}}$,
where $U_{\rm seed}= a T_{\rm seed}^4$ is the seed photon energy density generated on the surface of the disk,
and $a$ and $c$ are the radiation constant and the speed of light, respectively (see, e.g., \cite{mey00}).
From the energy balance within the disk, on the other hand,
we have $Q^{+}_{\rm{disk}}$~=~$Q^{-}_{\rm{disk}}$~=~$\sigma T_{\rm{seed}}^{4}$~=~$c U_{\rm{seed}}/4$,
where $Q^{+}_{\rm{disk}}$, $Q^{-}_{\rm{disk}}$, and $\sigma$ are
the heating rate, the cooling rate of the disk, and the Stefan-Boltzmann constant.
Suppose that most accretion energy liberated within the disk is transported to the corona by the magnetic field (e.g., \cite{haa91}).
Since about a half of the radiation scattered in the corona heats up the disk material,
the energy balance in the disk leads to $Q^{+}_{\rm{corona}}/2$~$\sim$~$Q^{+}_{\rm{disk}}$.
Finally we obtain $y$~$\sim$~1.
Note that the above calculations can apply only when a Comptonizing corona is optically thick ($\tau$~$\gtrsim$~1).

\medskip

\subsubsection{The high luminosity PL state --- possible scenario invoking a disk}

In the above scenario, the existence of a low-temperature optically-thick corona was assumed.
Inversely, we here consider another possibility;
that is, the spectral turnover in the high luminosity PL state may be given by direct disk emissions.
The origin of the turnover cannot be the standard disk,
since both spectra in XM2 and XM4 (namely the high luminosity PL state) were unaccountable by the MCD model (table~\ref{table5}).
Hence, we examined the slim disk emission by applying an extended MCD model (so-called the $p$-free model) with the additional parameter $p$,
which assumes a flatness of the temperature ($T$) profile of the disk; i.e., $T(r)$~$\propto$~$r^{-p}$, where $r$ is the radius of the disk.
The extended MCD model with $p$~=~0.75 is equal to the ``normal'' MCD model.
If the spectra are from the slim disk, it is expected that $p$ approaches from 0.75 to 0.5 as the X-ray luminosity increases (e.g., \cite{wata01}).
The same conclusion was derived, even if mass loss by a radiation-pressure driven outflow is taken into account \citep{take09}.
In fact, such a $p$--$L_{\rm{X}}$ variation was reported in a BHB \citep{kubo04a} and a ULX \citep{iso09} in the slim disk state. 
For IC\,342 X-1, the individual spectra can be reproduced by the extended MCD model (reduced-$\chi^2$~=~0.95 in XM2 and 0.92 in XM4),
but the derived $p$--$L_{\rm{X}}$ variation
($p$~=~0.54$\pm$0.01 and $L_{\rm{X}}$~=~1.0$\times 10^{40}$~erg~s$^{-1}$ in XM2 to
$p$~=~0.57$\pm$0.02 and $L_{\rm{X}}$~=~1.2$\times 10^{40}$~erg~s$^{-1}$ in XM4)
does not agree with the expected one.
We thus conclude that, at least, the IC\,342 X-1 spectra in the high luminosity PL state are from neither the standard disk nor the slim disk.

\medskip

\subsubsection{The low luminosity PL state}

Next, we discuss the low luminosity PL state, which is characterized by a PL tail extending up to at least 20~keV.
Since the low luminosity PL state also shows $y$~$\sim$~1, like the high luminosity PL state,
these two PL states can be described by a similar corona model with similar energy balance.
In other words, the distinction between the two PL states can be understood only by different coronal temperatures or coronal optical depth.

Two unresolved issues remain:
First, it is not clear which of $kT_{\rm{e}}$ or $\tau$ is the principle (controlling) parameter solely from the spectral fitting results.
If we stand on a scenario with the supercritical accretion rate, increase in $\tau$ is a cause of the electron temperature,
since the amount of outflow gas (i.e., $\tau$) can vary, depending on the disk luminosity \citep{kawa09}.
The observed behavior of the two PL states is consistent with their calculation.

\begin{figure}
\begin{center}
\FigureFile(72mm,57mm){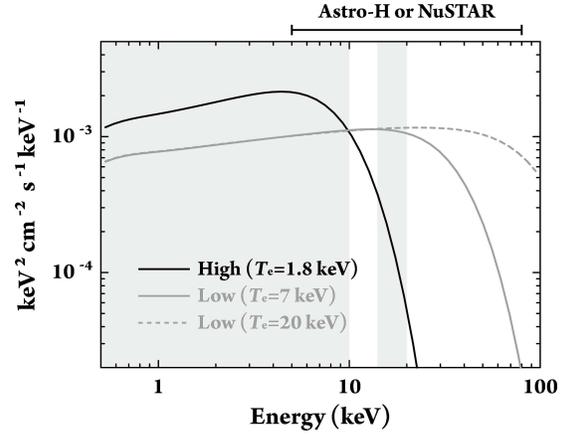}
\end{center}
\caption{Spectral energy distributions for the Comptonization model with parameters of each PL state.
The black solid line shows the high luminosity PL state with $kT_{\rm{e}}$~=~1.8~keV and $\tau$~=~8.5.
The gray lines show the low luminosity PL state
with $kT_{\rm{e}}$~=~7~keV and $\tau$~=~3.6 (solid) and $kT_{\rm{e}}$~=~20~keV and $\tau$~=~1.8 (dashed).
The seed photon temperature of both states is assumed to be 0.1~keV.
The range used in the present paper is shown with gray shading (0.5--10~keV and 14--20~keV).
The energy range of the Astro-H and the NuSTAR satellites is shown at top-right.
\label{fig10}}
\end{figure}

Second, the electron temperatures of the Comptonizing corona (or outflow, if it every exists) in the low luminosity PL state has not yet been precisely measured so far.
If the low luminosity PL state is somewhat similar to that of the VHS of Galactic BHBs, we expect higher electron temperatures, say, $kT_{\rm{e}}$~$\gtrsim$~20~keV.
Or alternatively, there may be no states like the VHS in ULXs, which is supported by the absence of large luminosity variations in ULXs.
In order to settle down this issue, more sensitive, higher energy ($>$20~keV) observations are required.
We illustrated the expected spectra of two representative cases in figure~\ref{fig10}.
In one case, there is a Comptonizing corona (or outflow) with a similar coronal electron temperature to that of the VHS ($kT_{\rm{e}}$~=~20~keV),
and hence the turnover energy is $\sim$60~keV.
While in the other case, the electron temperature is much less, $kT_{\rm{e}}$~=~7~keV, 
leading to a lower turnover energy is $\sim$20~keV.
Here, we fixed the optical depths to be $\tau$~=~1.8 and 3.6, respectively.
Such observations will be made possible by the launch of Astro-H \citep{koku08,taka10} and NuSTAR \citep{har05}.

\medskip

\subsubsection{Relation with the convex state}

Finally, we briefly mention the relation between the PL states and the convex state.
In IC\,342 X-1, the convex state has been observed on 1993 September with ASCA \citep{oka98,kubo01}.
Its spectral shape is apparently different from those in the PL states.
In fact, the spectrum in the convex state can be reproduced by the MCD model,
which does not reproduce the spectra in the PL states.
However, the 0.5--10~keV luminosity in the convex state ($\sim$1.3$\times 10^{40}$~erg~s$^{-1}$ assuming 3.3~Mpc)
is comparable to that in the high luminosity PL state.
That is, two distinct states appear at a similar luminosity.
These facts imply that the spectral state of ULXs (at least IC\,342 X-1) is not uniquely determined by only the luminosity (or the accretion rate)
but is determined in a more complicated fashion.
One possibility is a hysteretic relation known in Galactic BHBs, as we mentioned earlier ($\S$~5.2.1).
More frequent observations are required to elucidate complex spectral variations.

\section{Summary}

We have analyzed X-ray spectra of the unique ULX IC\,342 X-1
by using our two Suzaku data and archival data from multiple XMM-Newton, Chandra, and Swift observations.
X-1 clearly showed two distinctive sub-states:
(i) The low luminosity PL state appears in 70\% of all the present observations.
The X-ray luminosities are (4--6)$\times 10^{39}$~erg~s$^{-1}$ (0.5--10~keV), and the spectra have a PL tail extending up to at least 20~keV.
The coronal parameters derived from a Comptonization model are strongly coupled ($kT_{\rm{e}}$~$>$~2~keV and $\tau$~$<$~8).
(ii) The high luminosity PL state appears in 30\%.
The X-ray luminosities are (1.1--1.4)$\times 10^{40}$~erg~s$^{-1}$ (0.5--10~keV), and the spectra have a turnover at about 6~keV.
The coronal parameters are determined at $kT_{\rm{e}}$~$\sim$~1.8~keV and $\tau$~$\sim$~8.5.

Since the two PL states showed the similar Compton $y$-parameter of $y$~$\sim$~1,
we suggest that the two PL states are described by a similar corona model with similar energy balance.
The electron temperature and the optical depth respectively decreases and increases as an increasing X-ray luminosity.
This behavior of IC\,342 X-1 is similar
to that of some ULXs (especially Holmberg\,IX X-1) probably under the supercritical accretion situation,
or that of the Galactic BHB GRS\,1915$+$105 in the unique state with the supercritical accretion.
These facts imply a possibility that the two PL states of IC\,342 X-1 are formed by the supercritical accretion flows.

On the other hand, an unresolved issue for the relation between the high and low luminosity PL states remains.
More sensitive and higher energy observations achieved by the next generation X-ray satellites Astro-H and NuSTAR
would unveil the detailed physical condition of the low luminosity PL state.

\bigskip

We thank Kazuo Makishima for helpful comments for the paper.
This research has made use of data obtained from Data ARchives and Transmission System (DARTS),
provided by Center for Science-satellite Operation and Data Archives (C-SODA) at ISAS/JAXA,
and from the High Energy Astrophysics Science Archive Research Center (HEASARC), provided by NASA's Goddard Space Flight Center.
The Digitized Sky Survey was produced at the Space Telescope Science Institute under US Government grant NAG W-2166.
Images of this survey are based on photographic data obtained using the Oschin Schmidt Telescope on Palomar Mountain and the UK Schmidt Telescope.
The Second Digitized Sky Survey image used in this paper was produced with Montage,
which is an image mosaic service supported by the NASA Earth Sciences Technology Office Computing Technologies program.
Tessei Yoshida is supported by the Japan Society for the Promotion of Science Research Fellowship for Young Scientists.

\end{document}